\documentclass[pra,twocolumn]{revtex4}

\usepackage{amsmath,amsthm,graphicx,enumitem}   

\theoremstyle{remark}

\begin{document}

\newtheorem*{lem}{Lemma}
\newtheorem{prop}{Proposition}

\title{Quantifying nonlocality as a resource for device-independent quantum key distribution}
\author{S. Camalet}
\affiliation{Sorbonne Universit\'e, CNRS, Laboratoire de Physique 
Th\'eorique de la Mati\`ere Condens\'ee, LPTMC, F-75005, Paris, France}

\begin{abstract}
We introduce, for any bipartite Bell scenario, a measure that quantifies both the 
amount of nonlocality and the efficiency in device-independent quantum key 
distribution of a set of measurement outcomes probabilities. It is a proper 
measure of nonlocality as it vanishes when this set is Bell local and does not 
increase under the allowed transformations of the nonlocality resource theory. 
This device-independent key rate $R$ is defined by optimizing over a class of 
protocols, to generate the raw keys, in which each legitimate party does not use just 
one preselected measurement but randomly chooses at each round one among all 
the measurements at its disposal. A common and secret key can certainly be 
established when $R$ is positive but not when it is zero. For any continuous proper 
measure of nonlocality $N$, $R$ is tightly lower bounded by a nondecreasing 
function of $N$ that vanishes when $N$ does. There can thus be a threshold value 
for the amount of nonlocality as quantified by $N$ above which a secret key is surely 
achievable. A readily computable measure with such a threshold exists for two 
two-outcome measurements per legitimate party.
\end{abstract} 

\maketitle

\section{Introduction}

Using a secret sequence of characters, termed a key, for encryption and decryption,  
allows to transmit a message in an absolutely confidential way. The aim of quantum 
key distribution (QKD) studies is to examine whether two distant legitimate users, 
usually named Alice and Bob, can establish such a key in the presence of an 
eavesdropper, Eve, within the framework of quantum mechanics \cite{GRTZ}. To 
do so, Alice and Bob need at least to be able to generate, process and exchange 
random numbers and each to choose one out of several measurements to 
perform on quantum systems. The communication channel between them is public. 
Namely, any message sent over it becomes known to all parties. Moreover, Alice's 
and Bob's quantum systems in general share a global state with systems that Eve 
can manipulate as she wishes. On the other hand, Eve does not know which 
measurements Alice and Bob actually perform, the outcomes they get and the 
results  of their classical computations. The first security analyses of QKD schemes 
apply only to specific quantum systems Hilbert spaces and measurement operators 
on these spaces \cite{GRTZ,BB,E,BBM,LC,SP,KGR}. Consequently, a concrete 
implementation must follow the ideal model exactly. 

Device-independent QKD (DIQKD) protocols, on the contrary, do not require that 
Alice and Bob know anything about the sizes and states of the quantum systems 
and about the measurement devices \cite{ABGMPS,PABGMS}. They can only 
estimate the probabilities of the measurement outcomes. To establish a common 
and secret key, they first generate raw keys using measurent outcomes. These keys 
are not fully confidential and not completely identical to each other. Alice and 
Bob change them into the final key using random number generators, classical 
processors and the public channel. In Refs.\cite{ABGMPS,PABGMS}, only the 
so-called collective attacks, during the generation of the raw keys, are considered. 
Namely, it is assumed that Eve prepares a tripartite quantum system in the same 
state several times and that Alice's and Bob's possible measurements are the same 
each time. But the measurements may actually be performed on a global system 
which is not necessarily in a product state and the measurement devices may work 
differently from one round to another \cite{McK,MPA}. Furthermore, these 
apparatuses may have internal memories \cite{BCK,VV,ADFRV}. The security of a 
DIQKD protocol, for one execution, against these most general attacks follows 
from that against collective attacks \cite{ADFRV}. 

Device-dependent QKD is closely related to quantum entanglement. Some proposed 
protocols rely on entangled Alice's and Bob's quantum systems \cite{E,LC}. 
Moreover, the security of those known as prepare-and-measure protocols, for which 
such quantum correlations are absent, results from that of corresponding 
entanglement-based protocols \cite{BBM,SP,KGR}. Entanglement is a useful 
resource for many tasks and different measures of the entanglement of quantum 
states, appropriate for different tasks, have been introduced \cite{HHHH}. In more 
specific terms, entanglement theory is a resource theory. Entanglement cannot 
increase under local operations and classical communication and vanishes for separable 
states \cite{W,HHHH,VPRK,V,PRL1,PRL2}. Consequently, a proper measure of 
entanglement, called an entanglement monotone, is nonincreasing under these 
allowed transformations and is zero for separable states. The distillable key rate, 
defined for a given legitimate QKD users' state, satisfies these requirements 
and can be related to more familiar entanglement monotones \cite{HHHH,DW,KW}. 
In DIQKD, the necessary resource is not entanglement but Bell nonlocality 
\cite{E,ABGMPS,PABGMS,MY,BHK,McK,MPA,BCK,VV,ADFRV} whose relation to 
entanglement is not straightforward \cite{W,B,BCPSW,PRA}. A closely related issue 
which currently attracts much attention and in which Bell nonlocality is also essential 
is device-independent quantum random-number generation 
\cite{ADFRV,PAMBMMOHLMM,NBSP,ZKB,DF,BSC,MS}.

Bell nonlocality can also be formulated in terms of a resource theory \cite{DW,FWW,dV}. 
Proper measures of nonlocality must not increase under the corresponding allowed 
transformations, recalled in detail below, and vanish for Bell local sets of probabilities. 
We name these measures as nonlocality monotones. In this paper, we are interested in 
Bell nonlocality as a resource for DIQKD from this rigorous perspective. We introduce, 
for any numbers of choosable measurements and measurement outcomes, a measure 
$R$ which is both a nonlocality monotone and a DIQKD efficiency quantifier. This 
device-independent key rate is defined, under the assumption of collective attacks, 
by optimizing over a class of protocols, to generate the raw keys, which involve 
generating, processing and publicly exchanging random numbers and choosing at each 
round, for each legitimate user, one among all the possible measurements. Such a raw 
key protocol is a part of a full DIQKD protocol which also contains, for instance, an 
error correction part. A confidential key can surely be established when $R$ is positive 
but not when it is zero. The specific raw key protocols considered in the literature 
\cite{ABGMPS,PABGMS,McK,MPA,VV,ADFRV,BCK} belong to the class used here. We 
will see that a device-independent key rate defined for a single protocol can increase 
under the allowed transformations of the nonlocality resource theory. Since the rate 
$R$ is a nonlocality monotone, it does not decrease in going from a set of 
measurement outcomes probabilities to a more nonlocal one and vanishes for a Bell 
local set. Moreover, we show that, for any continuous nonlocality monotone $N$, $R$ 
is tightly lower bounded by a nondecreasing function of $N$ that vanishes when $N$ 
does. Thus, either this bound is trivial and nothing can be inferred from $N$ alone 
about the achievability of a secret key, or there is a threshold value for the amount of 
nonlocality as quantified by $N$ above which Alice and Bob are certain that such a key 
can be established, without needing to evaluate any other quantity.

The outline of the paper is as follows. In Sec.\ref{ABpo}, the allowed transformations 
of the nonlocality resource theory are recalled and the operations that the legitimate 
users can perform are specified, in terms of classical random variables. In 
Sec.\ref{Rkp}, we introduce the considered class of raw key protocols, which involve 
only these operations, and give the expression of the quantum state shared by the 
three parties at the end of such a protocol. In Sec.\ref{Dikr}, we define the 
device-independent key rate $R$ corresponding to this class of protocols and show 
that it is a nonlocality monotone. The case of a single protocol is also discussed in 
sec.\ref{Dikr}. In Sec.\ref{Cnm}, we consider the continuous nonlocality monotones, 
derive the above mentioned result, which follows from the fact that $R$ is a nonlocality 
monotone, and examine an example. Finally, in Sec.\ref{Soq}, we summarize our 
results and mention some open questions.

\section{Preliminaries}\label{P}

\subsection{Alice and Bob's possible operations}\label{ABpo}

\begin{figure}
\centering
\includegraphics[width=0.45\textwidth]{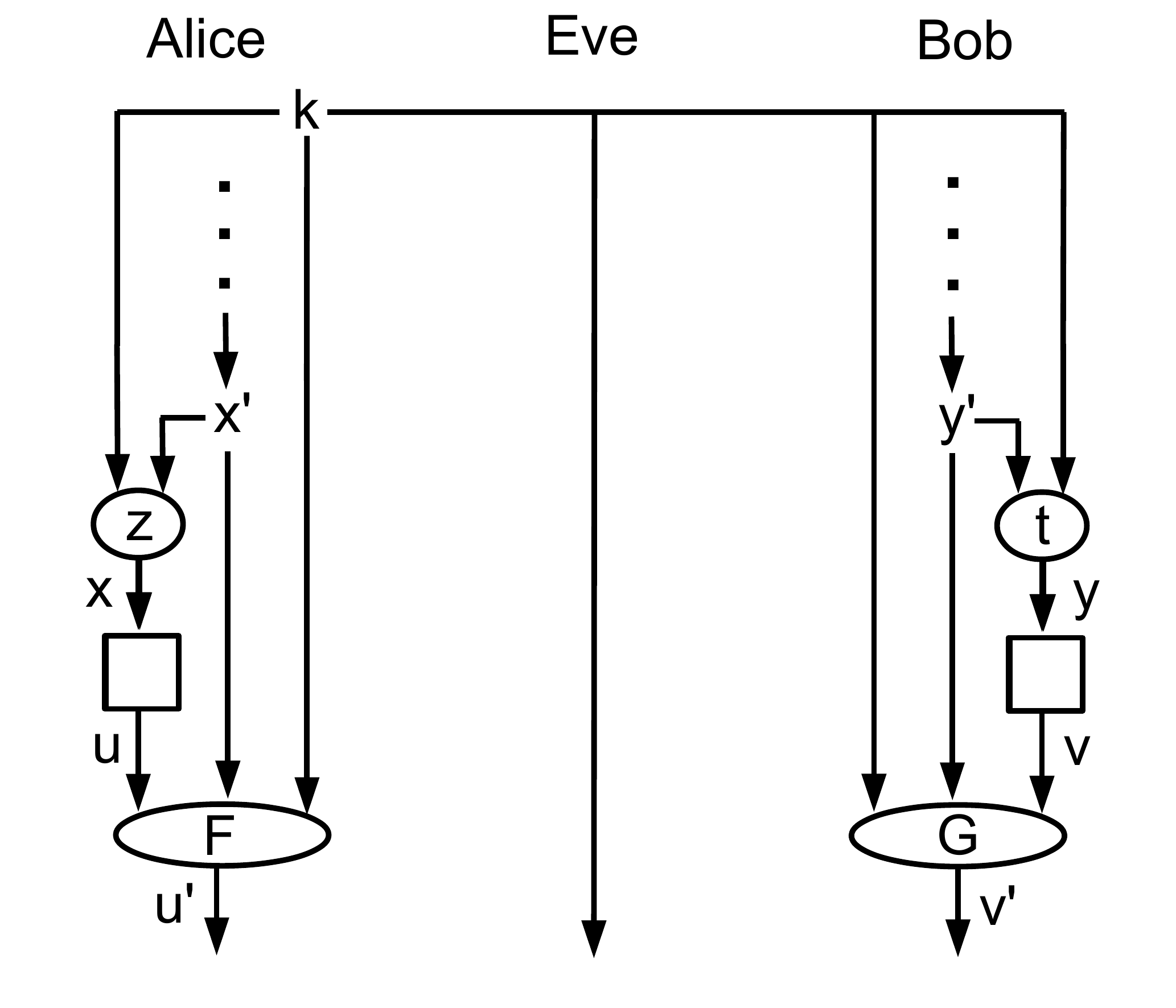}
\caption{Raw key protocol steps used in the proof of Proposition \ref{R}. The value 
$k$ is produced by a random generator. Eve eavesdrops $k$ sent by Alice over the 
public channel. The random numbers $x'$ and $y'$ are produced during classical 
steps, shown as dots, which can use the public channel and generate other random 
numbers, not shown. Classical computations, shown as ellipses, result in 
$x=z(x',k)$, $y=t(y',k)$, $u'=F_{x',k}(u)$ and $v'=G_{y',k}(v)$ with the 
functions given by the Lemma. The measurements performed by Alice and Bob on 
their quantum systems, shown as boxes, give the outputs $u$ and $v$ as functions 
of the inputs $x$ and $y$, respectively, according to the distributions \eqref{qpt}.}
\label{Fig}
\end{figure}

The following situation is considered throughout the paper. Alice, Bob and Eve 
initially share a quantum system in the state $\rho$. Alice (Bob) can choose one of 
$m$ ($n$) measurements to perform on her (his) subsystem with Hilbert space 
${\cal H}_A$ (${\cal H}_B$) which can always be assumed to be 
infinite-dimensional. The legitimate users know nothing about the quantum system, 
its state and the measurement devices. In more precise terms, Alice (Bob) can observe 
one of $m$ ($n$) classical random variables $A_x$ ($B_y$). Alice and Bob can only get 
information on the probability mass functions $P_{A_x,B_y}$, denoted $P_{x,y}$ 
in the following. The indices $x$ and $y$ are usually termed as inputs and the 
outcomes of the random variables $A_x$ and $B_y$ as outputs. Without loss of 
generality, it can be assumed that all variables $A_x$ ($B_y$) have the same set 
${\cal A}$ (${\cal B}$) of outputs by adding zero probability outcomes. These sets 
are referred to as alphabets. Obviously, $m$, $n$, ${\cal A}$ and ${\cal B}$ are 
known to Alice and Bob. The random variable $A_x$ ($B_y$) corresponds to a set 
of positive operators $M_{x,a}$ ($N_{y,b}$) such that 
$\sum_{a\in{\cal A}} M_{x,a}$ ($\sum_{b\in{\cal B}} N_{y,b}$) is the identity 
operator on ${\cal H}_A$ (${\cal H}_B$) and
\begin{equation}
P_{x,y}(a,b)
=\operatorname{tr}( \rho M_{x,a} \otimes N_{y,b} \otimes I_E) , 
\label{qpt}
\end{equation}
where $I_E$ is the identity operator on Eve's Hilbert space ${\cal H}_E$. 
A distribution tuple ${\boldsymbol P}=( P_{x,y}(a,b) )_{x,y,a,b}$ is said to be 
quantum if it can be written in this form with appropriate state and measurement 
operators.

In addition to the $A_x$ and $B_y$, the legitimate users can create random variables 
uncorrelated to the $A_x$ and $B_y$ and available at first only to one of them. Each 
can also compute new random variables from preexisting ones and use a classical 
public communication channel. Any message sent over this channel becomes known 
to the three parties and it is the only way to get a random variable from another 
party. It is not necessary to introduce explicitly additional variables for Eve since they 
can be taken into account by considering suitable system, state $\rho$ and 
measurements on her subsystem. Let us be more specific about how the $A_x$, that 
Alice cannot observe simultaneously, are employed. At some stage, and only at this 
stage, Alice uses one of the random variables at her disposal, say $X$, with alphabet 
in $\{1,\ldots,m\}$, to choose which $A_x$ to observe. More precisely, she generates 
$U$ according to $U=A_x$ when $X=x$. Bob uses the $B_y$ and $Y$ with alphabet 
in $\{1,\ldots, n\}$ in a similar way to produce the random variable $V$, see 
Fig.\ref{Fig}. We remark that the distribution tuble ${\bf P}$ is necessarily Bell local 
for simultaneously observable random variables $A_x$ and $B_y$ \cite{BCPSW,F}.

\subsection{Nonlocality resource theory}\label{Nrt}

We recall here the allowed transformations of the nonlocality resource theory 
\cite{DW,FWW,dV}. They can be performed using shared randomness and local 
probability transformations. More precisely, consider two distribution tuples 
${\boldsymbol P}$ and ${\boldsymbol P}'$ made up of no-signaling probabilities 
with the same output alphabets and numbers of inputs. The former is not less nonlocal 
than the latter if and only if
\begin{equation}
{\boldsymbol P}'=p_0 {\boldsymbol L} 
+ \sum_{k \ge 1} p_k {\cal T}_k({\boldsymbol P}) , 
\label{Bno}
\end{equation} 
where the probabilities $p_k$ obey $\sum_{k \ge 0} p_k=1$, ${\boldsymbol L}$ is 
a Bell local distribution tuple and ${\cal T}_k$ are compositions of input and output 
relabelings, output coarse grainings and input substitutions \cite{dV}. The nonlocality 
order is partial, i.e., some distribution tuples are not related by eq.\eqref{Bno}. 
We remark that a similar order can be defined for quantum states \cite{Bu}.

An input substitution acts on any distribution tuple ${\boldsymbol P}$ as follows. 
For some given $x$ and $x'$, every component $P_{x',y}(a,b)$ is replaced by 
$P_{x,y}(a,b)$ and the other ones remain unchanged, and similarly for given 
inputs $y$ and $y'$. An input relabeling consists in a permutation of the inputs 
$x$ or of the inputs $y$. It can be decomposed into input transpositions, i.e., 
transformations that swap every pair of components $P_{x,y}(a,b)$ and 
$P_{x',y}(a,b)$ for some given $x$ and $x'$ and leave the other ones unchanged, 
and similarly for given inputs $y$ and $y'$. The output transformations change 
only the probabilities $P_{x,y}(a,b)$ for a given $x$ or $y$. Under an output 
relabeling for $x$, every component ${P}_{x,y}(a,b)$ is replaced by 
${P}_{x,y}(\pi(a),b)$ where $\pi$ is a permutation on ${\cal A}$. An output coarse 
graining is characterized by an input, a subset of the corresponding output alphabet 
and an element of this subset, say $x$, ${\cal A}'$ and $a'$, respectively. Such a 
transformation changes every component ${P}_{x,y}(a,b)$ as follows. This probability 
becomes $\sum_{a''\in {\cal A}'} {P}_{x,y}(a'',b)$ for $a=a'$, is set to zero for 
$a \in {\cal A}'\setminus \{a'\}$ and remains the same for $a \notin {\cal A}'$.

A nonlocality monotone $N$ vanishes for Bell local distribution tuples and preserves 
the nonlocality order, i.e., $N({\boldsymbol P}')\le N({\boldsymbol P})$ for 
${\boldsymbol P}$ and ${\boldsymbol P}'$ related by eq.\eqref{Bno}. As a simple 
example, we consider, in the case of numbers of inputs $m=n=2$ and alphabets 
$\cal A$ and $\cal B$ consisting of two outputs, that can always be assumed to be 
$-1$ and $1$, the Clauser-Horne-Shimony-Holt inequality \cite{CHSH} 
violation 
\begin{equation}
\tilde N({\boldsymbol P})= \max\Big\{0,\max_\nu
\big|\sum_{x,y=1}^2 \nu(x,y) \langle A_x B_y \rangle\big|-2\Big\} .
\label{CHSH}
\end{equation}
In this expression, the maximum is 
taken over all the maps $\nu : \{ 1,2\}^2 \rightarrow \{ -1,1\}$ assuming the value 
$-1$ only once and $\langle C \rangle$ denotes the expectation of the random 
variable $C$. The measure \eqref{CHSH} vanishes for Bell local distribution tuples 
and only for them \cite{BCPSW}. To see that it preserves the nonlocality order, first 
note that it is a convex function of its argument. Moreover, the right side of 
eq.\eqref{CHSH} is not modified by an input relabeling. An output relabeling is 
equivalent to changing the sign of one of the random variables in eq.\eqref{CHSH}, 
and so also does not alter the value of $\tilde N$. An input substitution is the same 
as setting $A_1=A_2$ or $B_1=B_2$ in eq.\eqref{CHSH} which gives $\tilde N=0$. 
An output coarse graining is equivalent to replacing one of the random variables in 
eq.\eqref{CHSH} by $1$ which also leads to $\tilde N=0$.

\section{Raw key protocols}\label{Rkp}

To generate their raw keys, using the $A_x$ and $B_y$, Alice and Bob proceed as 
follows. First, they create some random variables and send some of them over the 
public channel. Then, they calculate new ones and subsequently produce the $U$ and 
$V$ as explained above. Finally, they generate $A$ and $B$ from all the available 
random variables. Alice's (Bob's) raw key is a sequence of independent realizations of 
$A$ ($B$). All the just mentioned classical random variables but $U$, $V$, $A$ and 
$B$ are quantum-mechanically described by the state
\begin{equation}
\tilde \rho = \sum_{{\boldsymbol x},{\boldsymbol y},{\boldsymbol e}} 
P_{{\boldsymbol X},{\boldsymbol Y},{\boldsymbol E}} 
({\boldsymbol x},{\boldsymbol y},{\boldsymbol e}) 
\Pi_{{\boldsymbol x},{\boldsymbol e}}^{Alice} 
\otimes \Pi_{{\boldsymbol y},{\boldsymbol e}}^{Bob} 
\otimes \Pi_{{\boldsymbol e}}^{Eve} ,
\label{tilderhot}
\end{equation}
where ${\boldsymbol E}$ is a tuple made up of the public ones and 
${\boldsymbol X}$ (${\boldsymbol Y}$) is made up of Alice's (Bob's) private ones. 
The choice random variable $X$ ($Y$) is a component of ${\boldsymbol X}$ 
(${\boldsymbol Y}$) or of ${\boldsymbol E}$. From their definitions, 
${\boldsymbol X}$ and ${\boldsymbol Y}$ are conditionally independent given 
${\boldsymbol E}$, i.e., $P_{{\boldsymbol X},{\boldsymbol Y},{\boldsymbol E}}
=P_{{\boldsymbol X}|{\boldsymbol E}}
P_{{\boldsymbol Y}|{\boldsymbol E}} P_{\boldsymbol E}$. 
The $\Pi_{{\boldsymbol x},{\boldsymbol e}}^{Alice}$ 
($\Pi_{{\boldsymbol y},{\boldsymbol e}}^{Bob}$, 
$\Pi_{{\boldsymbol e}}^{Eve}$) are rank-one projectors whose sum is the identity 
operator on a Hilbert space ${\cal H}'_A$ (${\cal H}'_B$, ${\cal H}'_E$). At the end 
of the raw key protocol, the three parties share the state
\begin{equation}
\rho_{\mathrm{rk}}= \sum_{a,b} \Pi_a \otimes \Pi_b \otimes
\operatorname{tr}_{{\cal H}_{AB}} 
(\rho' M_a \otimes N_b \otimes I'_E) , 
\label{rhof}
\end{equation}
where $\rho'=\tilde \rho \otimes \rho$, $I'_E$ is the identity operator on 
${\cal H}'_E\otimes{\cal H}_E$, $\Pi_a$ ($\Pi_b$) denotes mutually orthogonal 
rank-one projectors and $\operatorname{tr}_{{\cal H}_{AB}}$ the partial trace 
over the Hilbert space ${\cal H}_{AB}={\cal H}'_{A}\otimes{\cal H}_{A}
\otimes{\cal H}'_{B}\otimes{\cal H}_{B}$, see Appendix \ref{De3}. 
Observing $A$ and $B$ means performing the measurement with operators 
$\Pi_a \otimes \Pi_b \otimes I'_E$ on $\rho_{\mathrm{rk}}$. The positive 
operator $M_a$ reads
\begin{equation}
M_a= \sum_{{\boldsymbol x},{\boldsymbol e},u} 
P_{A|U,{\boldsymbol X},{\boldsymbol E}} 
(a|u,{\boldsymbol x},{\boldsymbol e})
\Pi_{{\boldsymbol x},{\boldsymbol e}}^{Alice} 
\otimes M_{x,u} ,
\label{Mat}
\end{equation} 
where $x$ corresponds to $X$ and the conditional probability mass function 
$P_{A|U,{\boldsymbol X},{\boldsymbol E}}$ is determined by the protocol. 
The operators $N_b$ are given by similar expressions. 

Equation \eqref{rhof} shows that the correlations between Alice, Bob and Eve at the 
end of the raw key protocol are formally identical to those obtained by performing 
the measurement with operators $M_a \otimes N_b \otimes I'_E$ on $\rho'$. The 
simple protocol in which Alice and Bob each just perform a given measurement, 
corresponding to the inputs, say $\xi$ and $\zeta$, is obviously one of those 
considered here. In this case, there is no public random variables, 
$P_X(x)=\delta_{x,\xi}$, $P_Y(y)=\delta_{y,\zeta}$, $A=U$ and $B=V$ and so 
the last term of eq.\eqref{rhof} simplifies to 
$\operatorname{tr}_{{\cal H}_A\otimes{\cal H}_B} 
(\rho M_{\xi,a} \otimes N_{\zeta,b} \otimes I_E)$. Strictly speaking, the above is 
only valid when Eve merely collects the public information in the course of the raw 
key protocol. However, any Eve's measurement during it can equivalently be taken 
into account as a measurement on $\rho_{\mathrm{rk}}$, see Appendix \ref{De3}. 
Thus, it can be considered as being performed amid the protocol generating 
the private key. 

\section{Device-independent key rate}\label{Dikr}

Let $\omega$ be a state of the form of eq.\eqref{rhof}, $l(\omega^{\otimes L})$ 
the length of the longest secret key that Alice and Bob can achieve when they share 
$\omega^{\otimes L}$ with Eve and $R'(\omega)$ the large $L$ limit of 
$l(\omega^{\otimes L})/L$ \cite{DW,M}. We define, for any quantum distribution 
tuple ${\boldsymbol P}$, the device-independent key rate 
\begin{equation}
R({\boldsymbol P}) = \inf_{\rho,(M_{x,a})_a,(N_{y,b})_b} 
\; \sup_{\mathrm{rkp}}\; R'(\rho_{\mathrm{rk}}) , \label{Rdef}
\end{equation}
where the supremum is taken over all the raw key protocols described above, the 
infimum is taken over all the $\rho$, $(M_{x,a})_a$ and $(N_{y,b})_b$ satisfying 
eq.\eqref{qpt} with ${\boldsymbol P}$ and $\rho_{\mathrm{rk}}$ is given by 
eqs.\eqref{tilderhot}-\eqref{Mat} with the probability mass functions of the 
protocol $\mathrm{rkp}$. The rate $R$ is nonnegative by construction. Whenever 
Alice's and Bob's random variables are described by the distributions $P_{x,y}$, they 
can establish, in the limit of large $L$, a secret key of length at least equal to 
$L R({\boldsymbol P})$ from raw keys of $L$ characters generated using an 
appropriate raw key protocol. In particular, a private key can surely be generated 
as soon as $R({\boldsymbol P})>0$. If $R({\boldsymbol P})=0$, there are states 
and measurement operators fulfilling eq.\eqref{qpt} with ${\boldsymbol P}$ for 
which a confidential key cannot be achieved.

\subsection{Rate for a single raw key protocol}\label{Rsrkp}

Let us first discuss the usual approach that considers only one specific protocol to 
generate the raw keys. The corresponding device-independent key rate is
\begin{equation}
R_0 ({\boldsymbol P}) = \inf_{\rho,(M_{x,a})_a,(N_{y,b})_b} 
R'(\rho_{\mathrm{rk}}) , \label{R0}
\end{equation}
where the infimum is taken over all the $\rho$, $(M_{x,a})_a$ and $(N_{y,b})_b$ 
satisfying eq.\eqref{qpt} with ${\boldsymbol P}$ and $\rho_{\mathrm{rk}}$ is given 
by eqs.\eqref{tilderhot}-\eqref{Mat} with the distributions of the particular protocol 
employed. As an example, assume that $m=3$, $n=2$, ${\cal A}={\cal B}=\{-1,1\}$, 
$A=EA_3$ and $B=EB_1$ where $E$ is an equally distributed public random variable 
with alphabet ${\cal A}$ \cite{ABGMPS}. The rate $R'(\rho_{\mathrm{rk}})$ is not 
larger than the mutual information $I$ between $A$ and $B$ \cite{M} which can be 
expressed in terms of $P_{3,1}$ with 
$P_{A,B}(a,b)=(P_{3,1}(a,b)+P_{3,1}(-a,-b))/2$. Provided that the Bell expression 
$S= \sum_{x,y=1}^2 (-1)^{(x-1)(y-1)} \langle A_x B_y \rangle$ is larger than its 
Bell local maximum of $2$, $R'(\rho_{\mathrm{rk}})$ is not lower than 
$I-h(1/2+\sqrt{S^2/4-1}/2)$ where $h$ is the binary entropy function 
\cite{ABGMPS}. Since these two bounds depend only on ${\boldsymbol P}$, they are 
also bounds for $ R_0({\boldsymbol P})$ given by eq.\eqref{R0}.

Let ${\bf P}$ and ${\bf P}'$ be the distribution tuples defined by 
$P_{1,1}(a,b)=P_{2,1}(a,b)=(1+ab\cos\theta)/4$, 
$P_{1,2}(a,b)=(1+ab\sin\theta)/4$, $P_{2,2}(a,b)=(1-ab\sin\theta)/4$, 
$P_{3,1}=P_{3,2}=1/4$ and $P'_{x,y}=P_{z(x),y}$ where $\theta$ is any real 
number, $z(1)=z(3)=1$ and $z(2)=2$. They are quantum since ${\boldsymbol P}$ 
(${\boldsymbol P}'$) can, for instance, be written as 
$P_{x,y}(a,b)=\langle \psi | \Pi_{x,a}^A \otimes \Pi_{y,b}^B |\psi \rangle$
with the two-qubit maximally entangled state 
$|\psi \rangle=(|+\rangle\otimes|+\rangle+|-\rangle\otimes|-\rangle)/\sqrt{2}$ 
where $|\pm \rangle$ are orthonormal states of the Hilbert space ${\cal H}_2$ of 
dimension 2 and the projective measurement operators 
$\Pi^A_{x,1}=|x \rangle_A {_A\langle} x |$ 
($\Pi^A_{x,1}=|z(x) \rangle_A {_A \langle} z(x) |$), 
$\Pi^B_{y,1}=| y \rangle_B {_B\langle} y |$, 
$\Pi^B_{x,-1}=I_2-\Pi_{x,1}$ and $\Pi_{y,-1}=I_2-\Pi_{y,1}$ where $I_2$ is 
the identity operator on ${\cal H}_2$, 
$|1 \rangle_A=\cos \phi|+\rangle+\sin \phi|-\rangle$, 
$|2 \rangle_A=\cos \phi|+\rangle-\sin \phi|-\rangle$, 
$|3 \rangle_A=(|+\rangle+i|-\rangle)/\sqrt{2}$, $|1 \rangle_B=|+ \rangle$ and 
$|2 \rangle_B=(|+\rangle+|-\rangle)/\sqrt{2}$ with $\phi=\theta/2$. For 
${\bf P}$ and ${\bf P}'$, $S=2\sqrt{2}\cos(\theta - \pi/4)$ increases from its Bell local 
maximum of $2$ to its quantum maximum of $2\sqrt{2}$ as $\theta$ varies from $0$ 
to $\pi/4$. Using the bounds mentioned above, one finds $R_0({\boldsymbol P})=0$ 
for any value of $\theta$, as $A$ and $B$ are uncorrelated for ${\boldsymbol P}$, and 
$R_0({\boldsymbol P}') \ge 1 - h(1/2+\cos\theta/2)-h(1/2+\sqrt{\sin(2\theta)}/2)$ 
for $\theta \in [0,\pi/2]$, and hence $R_0({\boldsymbol P}')>R_0({\boldsymbol P})$ 
for $\theta \in (0,1.032]$. On the other hand, ${\boldsymbol P}$ is not less nonlocal 
than ${\boldsymbol P}'$ since it can be transformed into ${\boldsymbol P}'$ by an 
input substitution. Consequently, $R_0$ is not a nonlocality monotone. 

In device-independent quantum random-number generation, a raw string is first 
generated following a given procedure. The corresponding appropriate rate can be 
lower bounded in terms of one or several Bell expressions 
\cite{PAMBMMOHLMM,NBSP,ZKB,DF,BSC,MS}, which shows a clear influence 
of nonlocality already for a single raw string protocol. In DIQKD, similar bounds 
on Eve's information on Alice's or Bob's raw key can be derived for a given raw key 
protocol \cite{ABGMPS,PABGMS,McK,MPA,VV,ADFRV}. But, in order to establish 
a common secret key, correlations between the two raw keys are also essential. 
In the example discussed above, for instance, the mutual information between the 
outcomes of $A$ (or $B$) and of any Eve's measurement is lower than 
$h(1/2+\sqrt{S^2/4-1}/2)$ which is zero at $S=2\sqrt{2}$ \cite{ABGMPS}. 
However, this does not ensure a nonzero rate $R_0$ since, for any value of $S$ in 
$[2,2\sqrt{2}]$, there are distributions tuples ${\bf P}$ for which $A$ and $B$ 
are uncorrelated and hence $R_0({\bf P})$ vanishes. 

\subsection{Main result}\label{Mr}

It can be proved that the rate \eqref{Rdef} preserves the Bell nonlocality order 
using the Lemma below. 
\begin{lem} Let ${\boldsymbol P}$ and ${\boldsymbol P}'$ be two distribution 
tuples with output alphabets ${\cal A}$ and ${\cal B}$ and numbers $m$ and $n$ 
of inputs such that ${\boldsymbol P}$ is not less nonlocal than ${\boldsymbol P}'$ 
and $A_x$ and $B_y$ be random variables such that $P_{A_x,B_y}=P_{x,y}$. 

There are a random integer $K$ and tuples 
${\boldsymbol {\tilde A}}=(\tilde A_x)_{x=1}^m$ 
and ${\boldsymbol {\tilde B}}=(\tilde B_y)_{y=1}^n$ with a joint probability mass 
function of the form $P_K P_{{\boldsymbol {\tilde A}},{\boldsymbol {\tilde B}}}$, 
where $\tilde A_x$ and $\tilde B_y$ have alphabets ${\cal A}$ and ${\cal B}$, 
respectively, self-maps $F_{x,k}$ and $G_{y,k}$ and inputs $z(x,k)$ and $t(y,k)$ 
such that the distributions of 
\begin{multline}
(A'_x,B'_y) = (\tilde A_x,\tilde B_y) \;\mathrm{if}\; K=0 \label{lem} \\
= \big(F_{x,k}(A_{z(x,k)}),G_{y,k}(B_{t(y,k)})\big) \;\mathrm{if}\; K=k \ge 1 , 
\end{multline}
are given by $P_{A'_x,B'_y}=P'_{x,y}$.

If ${\boldsymbol P}$ is quantum then also is ${\boldsymbol P}'$.
\end{lem}
The proof of the Lemma is given in Appendix \ref{PL}. This Lemma ensures that, from 
given $A_x$ and $B_y$ with distributions $P_{x,y}$, the legitimate users have 
effectively access to random variables characterized by any distribution tuple 
${\boldsymbol P}'$ not more nonlocal than ${\boldsymbol P}$, by proceeding as 
follows. Alice creates the corresponding $K$, $\tilde A_x$ and $\tilde B_y$ and sends 
$K$ and the $\tilde B_y$ to Bob. Then, Alice and Bob perform any classical operations 
they want, possibly using the public channel, that produce some random variables, 
including $X'$ and $Y'$ with alphabets in $\{1,\ldots,m\}$ and $\{1,\ldots,n\}$, 
respectively. After these classical steps, Alice generates, in sequence, the random 
variables $X$ which is $X'$ if $K=0$ and $z(X',k)$ if $K=k \ge 1$, $U$ according to 
$U=A_{x}$ when $X=x$ and, finally, $U'$ which is $\tilde A_{x'}$ if $(K,X')=(0,x')$ 
and $F_{x',k} (U)$ if $(K,X')=(k,x')$ with $k \ge 1$. Bob does similar operations using 
$Y'$, $K$, the $B_y$ and the $\tilde B_y$, see Fig.\ref{Fig}. In this Figure, it is 
assumed, to simplify, that $K \ge 1$. The above produces the same $U'$ and $V'$ as 
$U'=A'_{x'}$ when $X'=x'$ and $V'=B'_{y'}$ when $Y'=y'$ where $A'_{x'}$ and 
$B'_{y'}$ are given by eq.\eqref{lem}. Note that Eve gets $K$ and the $\tilde B_y$ 
which are sent over the public channel. 

Using the above Lemma, the following can be shown, see Appendix \ref{PP1}.
\begin{prop}\label{R}
The function $R$ given by eq.\eqref{Rdef} has the properties:
\begin{enumerate}[label=(\roman*)]
\item $R$ preserves the nonlocality order.
\item $R$ vanishes for Bell local distribution tuples.
\end{enumerate}
\end{prop}
A function fulfilling these two requirements is a nonlocality monotone, i.e., a proper 
measure of Bell nonlocality \cite{dV}.  The above Proposition implies that the 
device-independent key rate \eqref{Rdef} is a nonlocality monotone. Property (ii) 
can be seen as a consequence of the fact that a secret key cannot always be 
established and of (i) as follows. Any distribution tuple is not less nonlocal than any 
Bell local one. Thus, due to property (i), $R$ assumes its minimum value for Bell local 
distribution tuples. If this minimum were nonzero then a secret key could be produced 
in any case. Proposition \ref{R} shows that a private key can surely be generated for 
any distribution tuple not less nonlocal than a given one  for which this is possible. 
Besides, a confidential key can be established with certainty only for nonlocal 
distribution tuples. Proposition \ref{R} does not ensure that the converse holds. There 
may be nonlocal distribution tuples ${\boldsymbol P}$ such that a private key cannot 
be achieved for some states and measurement operators fulfilling eq.\eqref{qpt} with 
${\boldsymbol P}$.

\section{Continuous nonlocality monotones}\label{Cnm}

According to the above Proposition and definition \eqref{Rdef}, the device-independent 
key rate $R$ quantifies both the efficiency in secret key generation and the amount of 
Bell nonlocality of a distribution tuple. However, it is not straightforward to evaluate. 
Moreover, one may prefer a measure that provably vanishes only for Bell local 
distribution tuples. It is then of interest to consider other nonlocality monotones. For 
that purpose, we use the following result, shown in Appendix \ref{PP2}. We remark 
that the set of quantum distribution tuples depends on the dimensions of the 
considered Hilbert spaces \cite{PV}.
\begin{prop}\label{prop2}
Let ${\cal Q}$ and ${\cal L}$ be, respectively, the sets of quantum and Bell local 
distribution tuples with given output alphabets and numbers of inputs, for given Hilbert 
spaces dimensions.

For any nonlocality monotone $M$ on ${\cal Q}$ and nonnegative continuous 
function $N$ on ${\cal Q}$ which vanishes on ${\cal L}$, there is a nondecreasing 
function $f$ on $J=[0,N_{\mathrm{sup}})$, where $N_{\mathrm{sup}}$ is the 
supremum of $N$ on ${\cal Q}$, such that $f(0)=0$, $f \circ N \le M$ and, for any 
$s \in J$ and $\epsilon>0$, there is $\boldsymbol P \in {\cal Q}$ for which 
$N(\boldsymbol P)=s$ and $M(\boldsymbol P)<f(s)+\epsilon$.
\end{prop}
Whenever Alice's and Bob's random variables are described by the distributions 
$P_{x,y}$, they can generate a secret key with a rate not lower than 
$f \circ N({\boldsymbol P})$ where $N$ is any continuous nonlocality monotone and 
$f$ is given by the above Proposition with $N$ and $M=R$. This remains valid if $f$ 
is replaced by other nondecreasing functions but $f$ is the greatest one. If 
$f \circ N({\boldsymbol P})=0$, there exists, for any $\epsilon>0$, a quantum 
distribution tuple ${\boldsymbol P}'$ such that 
$N({\boldsymbol P}')=N({\boldsymbol P})$ and $R({\boldsymbol P}')<\epsilon$. 
So, in this case, nothing can be inferred from the value $N({\boldsymbol P})$ 
regarding the possibility of establishing a confidential key. On the contrary, 
$f \circ N({\boldsymbol P})>0$ ensures that a private key can be generated. This 
condition can be rewritten as $N({\boldsymbol P}) > N^*$ where 
$N^*=\sup\{s \in J : f(s)=0\}$ is set only by the measure $N$. By definition, 
$N^*$ is not larger than $N_{\mathrm{sup}}$. If $N^*=N_{\mathrm{sup}}$, 
$f \circ N=0$ and it cannot be determined whether a secret key can be established 
by evaluating only $N$. By contrast, for a measure $N$ such that 
$N^*<N_{\mathrm{sup}}$, $N^*$ is a threshold value above which a private key can 
be achieved with certainty. Proposition \ref{prop2} does not require that $N$ is a 
nonlocality monotone but only that it is continuous and vanishes for Bell local tuples. 
For instance, $N$ can be defined from a Bell inequality. Proposition \ref{prop2} applies 
to such measures though they are not nonlocality monotones in general \cite{dV}. 

As an example, assume that $m=n=2$ and ${\cal A}={\cal B}=\{-1,1\}$. In this case, 
the measure $\tilde N$, given by eq.\eqref{CHSH}, is a nonlocality monotone. As is 
well known, the set of the possible values of $\tilde N$ is the interval 
$[0,2(\sqrt{2}-1)]$ \cite{C}. A nondecreasing function $g$ such that 
$R \ge g \circ \tilde N$ can be found, see Appendix \ref{Ubt}. It results from its 
expression that there is a threshold value $\tilde N^* \le 0.652 < 2(\sqrt{2}-1)$ for 
the nonlocality monotone \eqref{CHSH} above which a private key can surely be 
generated. The existence of $\tilde N^*$ can be seen as follows. For any 
${\boldsymbol P}$ such that $\tilde N({\boldsymbol P})>0$, there are inputs $\xi$ 
and $\zeta$ for which $|\langle A_\xi B_\zeta \rangle| > 0$. Consider a raw key 
protocol generating $A=E A_\xi$ and $B=E B_\zeta$ where $E$ is an equally 
distributed public random variable with alphabet $\cal A$. The resulting rate 
$R'(\rho_{\mathrm{rk}})$ in eq.\eqref{Rdef}, and hence $R({\boldsymbol P})$, is 
not lower than $I-  r \circ \tilde N({\boldsymbol P})$ where $r$ is a continuous 
nonincreasing function with $r(2\sqrt{2}-2)=0$, given by 
$r(s)=h(1/2+(s+s^2/4)^{1/2}/2)$, and $I=1 - h(1/2+|\langle A B \rangle|/2)$ is 
the mutual information between $A$ and $B$ that depends only on 
$|\langle A B \rangle|=|\langle A_\xi B_\zeta \rangle|$ and is hence strictly positive 
for $\tilde N({\boldsymbol P})>0$ \cite{ABGMPS}. The function $g$ can be obtained 
by noting that there are $\xi$ and $\zeta$ such that 
$|\langle A B \rangle| \ge 1/2+\tilde N({\boldsymbol P})/4$. Other raw key protocols 
are used in Appendix \ref{Ubt}.

\section{Summary and open questions}\label{Soq}

In summary, a device-independent key rate has been defined by optimizing over a class 
of raw key protocols and shown to be a nonlocality monotone. Moreover, it has been 
proved that there are only two possibilities for any continuous nonlocality monotone. 
Either it can never be decided whether a secret key can be established by evaluating 
only this measure, or there is a threshold value for it above which this is surely 
achievable. A readily computable nonlocality monotone with such a threshold exists for 
two two-outcome measurements per legitimate user. The defined device-independent 
key rate may vanish for some nonlocal sets of probabilities. Were this not to be the 
case, Bell nonlocality would be a necessary and sufficient condition for DIQKD with raw 
key protocols. This may be correct only for some numbers of choosable measurements 
and measurement outcomes. Related to this issue, it would be interesting to improve 
the upper bound on the threshold value of the aforementioned particular nonlocality 
monotone. Since this measure vanishes only for Bell local sets of probabilities, a 
threshold value of zero would prove the above mentioned equivalence in this case. 
The answers to these open questions may depend on the considered class of raw key 
protocols. It can be further enlarged, for instance, by dropping the assumption made 
here that no information is exchanged after the measurements.

\begin{appendix}

\section{Derivation of equation \ref{rhof}}\label{De3}

To simplify, the random variables transmitted over the public channel 
at the same stage of the raw key protocol are here grouped into one. 
After receiving the value $e_1$ of the first one $E_1$, Eve performs 
a measurement on her subsystem. This generates a random variable 
$E'_1$ and $\rho$ is changed into 
$\Lambda_{e_1,e'_1}(\rho)/p_{e_1,e'_1}$ when $E'_1=e'_1$ 
where $p_{e_1,e'_1}=\operatorname{tr} \Lambda_{e_1,e'_1}(\rho)$. 
The Kraus operators of the quantum operation $\Lambda_{e_1,e'_1}$ 
are of the form $I_A\otimes I_B \otimes K_{e_1,e'_1,i}$ where
$I_A$ ($I_B$) is the identity operator on ${\cal H}_A$ (${\cal H}_B$). 
The probabilities $p_{e_1,e'_1}$ satisfy $\sum_{e'_1} p_{e_1,e'_1}=1$. 
A deterministic operation, e.g, the identity operation, is a measurement 
with a single outcome $e'_1$. Moreover, sequential measurements can 
be considered as a single one with a properly defined $E'_1$. The set 
of the values $e'_1$ may depend on $e_1$. However, it can be assumed, 
without loss of generality, that it does not, by adding zero probability 
outcomes, and hence that there is a unique $E'_1$ with 
$P_{E'_1|E_1}(e'_1|e_1)=p_{e_1,e'_1}$. Repeating these arguments 
for all components of ${\boldsymbol E}$ leads to the random tuple 
${\boldsymbol E'}$, conditional distribution 
$P_{{\boldsymbol E'}|{\boldsymbol E}}$ and quantum operations 
$\Lambda_{{\boldsymbol e},{\boldsymbol e'}}$ with Kraus operators 
$I_A\otimes I_B \otimes K_{{\boldsymbol e},{\boldsymbol e'},i}$. 

The probability mass function of ${\boldsymbol X}$, ${\boldsymbol Y}$, 
${\boldsymbol E}$, ${\boldsymbol E'}$, $U$ and $V$ is 
$P_{{\boldsymbol X},{\boldsymbol Y},{\boldsymbol E}}
P_{{\boldsymbol E'}|{\boldsymbol E}}
P_{U,V|{\boldsymbol X},{\boldsymbol Y},{\boldsymbol E},{\boldsymbol E'}}$ 
where the last conditional distribution is given by 
$$P(u,v|{\boldsymbol x},{\boldsymbol y},{\boldsymbol e},{\boldsymbol e'})
=\operatorname{tr}( \Lambda_{{\boldsymbol e},{\boldsymbol e'}}(\rho) 
M_{x,u} \otimes N_{y,v} \otimes I_E'')/P
({\boldsymbol e'}|{\boldsymbol e}) ,  $$
with the appropriate identity operator $I''_E$, and omitting the subscripts for 
the distributions. For given values of these random variables, Eve's state 
is proportionnal to $\Lambda'_{{\boldsymbol e},{\boldsymbol e'}} 
(\operatorname{tr}_{{\cal H}_A\otimes{\cal H}_B}
( \rho M_{x,u} \otimes N_{y,v} \otimes I_E))$ 
where $\Lambda'_{{\boldsymbol e},{\boldsymbol e'}}$ is the quantum 
operation with Kraus operators $K_{{\boldsymbol e},{\boldsymbol e'},i}$. 
The marginal distribution $P_{A,B,{\boldsymbol E},{\boldsymbol E'}}$ 
directly follows with the conditional distributions 
$P_{A|U,{\boldsymbol X},{\boldsymbol E}}$ and 
$P_{B|V,{\boldsymbol Y},{\boldsymbol E}}$ of the protocol. 
Using $P_{{\boldsymbol X},{\boldsymbol Y},U,V
|A,B,{\boldsymbol E},{\boldsymbol E'}}$, 
one finds that Eve's state for $A=a$, $B=b$, 
${\boldsymbol E}={\boldsymbol e}$ and 
${\boldsymbol E'}={\boldsymbol e'}$ is
\begin{multline}
\omega_{a,b,{\boldsymbol e},{\boldsymbol e'}} 
= \Lambda_{{\boldsymbol e'}} \Big( \Pi_{{\boldsymbol e}}^{Eve}
\otimes \sum_{{\boldsymbol x},{\boldsymbol y},u,v}
P(a|u,{\boldsymbol x},{\boldsymbol e})
P(b|v,{\boldsymbol y},{\boldsymbol e}) \\
\times P({\boldsymbol x},{\boldsymbol y},{\boldsymbol e})
\operatorname{tr}_{{\cal H}_A\otimes {\cal H}_B}
( \rho M_{x,u} \otimes N_{y,v} \otimes I_E)\Big)
/P(a,b,{\boldsymbol e},{\boldsymbol e'}) ,\nonumber
\end{multline}
where $\Pi_{{\boldsymbol e}}^{Eve}
=|{\boldsymbol e}\rangle \langle {\boldsymbol e} |$ 
and $\Lambda_{{\boldsymbol e'}}$ is the quantum operation with 
Kraus operators 
$\langle {\boldsymbol e} | \otimes K_{{\boldsymbol e},{\boldsymbol e'},i}$. 
Performing the measurement with operators 
$\Pi_a\otimes\Pi_b\otimes\Pi_{{\boldsymbol e}}^{Eve}\otimes I_E$ and 
then that characterized by the $\Lambda_{{\boldsymbol e'}}$ on the state 
given by eq.\eqref{rhof} leads to the same distribution 
$P_{A,B,{\boldsymbol E},{\boldsymbol E'}}$ and 
Eve's states $\omega_{a,b,{\boldsymbol e},{\boldsymbol e'}}$. 

\section{Proof of the Lemma}\label{PL}

The tuples ${\boldsymbol P}$ and ${\boldsymbol P}'$ are related by equation 
\eqref{Bno}. The Bell local distribution tuple ${\boldsymbol L}$ can be written 
as ${\boldsymbol L}=\sum_{{\boldsymbol a},{\boldsymbol b}} 
q_{{\boldsymbol a},{\boldsymbol b}} 
{\boldsymbol D}_{{\boldsymbol a},{\boldsymbol b}} $ where the sum 
runs over all the ${\boldsymbol a}=(a_x)_x \in {\cal A}^n$ and 
${\boldsymbol b}=(b_y)_y \in {\cal B}^m$, the probabilities 
$q_{{\boldsymbol a},{\boldsymbol b}} $ sum to unity and the only 
nonvanishing components of 
${\boldsymbol D}_{{\boldsymbol a},{\boldsymbol b}} $, 
for the inputs $x$ and $y$, are those corresponding to the outputs 
$a=a_x$ and $b=b_y$ \cite{BCPSW}. Consider random tuples 
${\boldsymbol {\tilde A}}=(\tilde A_x)_{x=1}^m$ and 
${\boldsymbol {\tilde B}}=(\tilde B_y)_{y=1}^n$ where $\tilde A_x$ and 
$\tilde B_y$ have alphabets ${\cal A}$ and ${\cal B}$, respectively, such 
that $P_{{\boldsymbol {\tilde A}},{\boldsymbol {\tilde B}}}  
({\boldsymbol a},{\boldsymbol b})=q_{{\boldsymbol a},{\boldsymbol b}}$. 
The components of ${\boldsymbol L}$ are equal to the marginal 
probabilities $P_{\tilde A_x,\tilde B_y}(a,b)$.

We denote an input substitution for $x$ and $x'$ as ${\cal I}^{(x,x')}$ and 
an input transposition for $x$ and $x'$, described in the main text, 
as ${\cal I}^{\{x,x'\}}$, and similarly for given inputs $y$ and $y'$. The input 
transformations satisfy ${\cal I}^{\{y,y'\}} \circ {\cal I}^{\{x,x'\}}
={\cal I}^{\{x,x'\}} \circ {\cal I}^{\{y,y'\}}$ and similar commutation relations with 
one or both transformations replaced by an input substitution. We denote the output 
transformations as ${\cal O}^x$ and ${\cal O}^y$. They obey 
${\cal O}^y \circ {\cal O}^x={\cal O}^x \circ {\cal O}^y$. We have also 
${\cal I}^{(x,x')} \circ {\cal O}^z={\cal O}^z \circ {\cal I}^{(x,x')}$ for 
$z \ne x,x'$, ${\cal I}^{(x,x')} \circ {\cal O}^{x'}={\cal I}^{(x,x')}$, 
${\cal I}^{(x,x')} \circ {\cal O}^{x}
={\cal O}^x \circ {\cal O}^{x'} \circ {\cal I}^{(x,x')}$, 
${\cal I}^{\{x,x'\}} \circ {\cal O}^z={\cal O}^z \circ {\cal I}^{\{x,x'\}}$ for 
$z \ne x,x'$, ${\cal I}^{\{x,x'\}} \circ {\cal O}^x
={\cal O}^{x'} \circ {\cal I}^{\{x,x'\}}$ and 
similar relations with the inputs $x$ and $x'$ replaced by inputs $y$ 
and $y'$. Consequently, any transformation ${\cal T}_k$ appearing in 
eq.\eqref{Bno} can be written as 
${\cal T}_k={\cal O}_k^A\circ{\cal O}_k^B
\circ{\cal I}^A_k\circ{\cal I}^B_k$ where ${\cal O}^A_k$ 
(${\cal O}^B_k$) consists of output transformations for given inputs $x$ ($y$) 
and ${\cal I}^A_k$ (${\cal I}^B_k$) of transformations on the inputs $x$ ($y$). 
Moreover, the component of 
${\cal I}^A_k\circ{\cal I}^B_k({\boldsymbol P})$ for the inputs $x$ and $y$ 
and outputs $a$ and $b$ can be expressed as $P_{A_{z(x,k)},B_{t(y,k)}}(a,b)$ 
with $z(x,k)$ and $t(y,k)$ determined by ${\cal I}^A_k$ and ${\cal I}^B_k$, 
respectively. 

Let ${\boldsymbol {\hat P}}$ be any distribution tuple with alphabets 
${\cal A}$ and ${\cal B}$ and $C_x$ and $D_y$ random variables such 
that $P_{C_x,D_y}={\hat P}_{x,y}$. An output relabeling for $x$ acts on 
${\boldsymbol {\hat P}}$ as follows. Every component ${\hat P}_{x,y}(a,b)$ 
is replaced by ${\hat P}_{x,y}(\pi(a),b)=P_{\pi^{-1}(C_x),D_y}(a,b)$ 
where $\pi$ is a permutation on ${\cal A}$ and $\pi^{-1}$ is its inverse 
and the other ones remain unchanged, and similarly for a given input $y$. 
Under an output coarse graining characterized by $x$, 
${\cal A}' \subset {\cal A}$ and $a' \in {\cal A}'$, every component 
${\hat P}_{x,y}(a,b)$ becomes $\sum_{a''\in {\cal A}'} {\hat P}_{x,y}(a'',b)$ 
for $a=a'$, vanishes for 
$a \in {\cal A}'\setminus \{a'\}$, 
and does not change for $a \notin {\cal A}'$ and 
the other ones remain the same, and similarly for given $y$, 
${\cal B}' \subset {\cal B}$ and $b' \in {\cal B}'$. The components for 
$x$ of the resulting distribution tuple can be written as 
$P_{F(C_x),D_y}(a,b)$ where the self-map $F$ on ${\cal A}$ is given 
by $F(a)=a$ for $a \notin {\cal A}'$ and $F(a)=a'$ for $a \in {\cal A}'$. 
Consequently, the component for the inputs $x$ and $y$ and outputs 
$a$ and $b$ of ${\cal T}_k({\boldsymbol P})$ in eq.\eqref{Bno} can be 
expressed as $P_{F_{x,k}(A_{z(x,k)}),G_{y,k}(B_{t(y,k)})}(a,b)$ with 
self-maps $F_{x,k}$ and $G_{y,k}$ on ${\cal A}$ and ${\cal B}$, 
respectively, determined by ${\cal O}^A_k$ and ${\cal O}^B_k$, 
respectively. 

The above shows that $$P'_{x,y}=p_0 P_{\tilde A_x,\tilde B_y} 
+ \sum_{k \ge 1} p_k P_{F_{x,k}(A_{z(x,k)}),G_{y,k}(B_{t(y,k)})} .$$ 
It remains to introduce a random non-negative integer $K$ with distribution 
$P_K(k)=p_k$ and the random variables $A'_x$ and $B'_y$ given 
in the Lemma. The probability mass function of $K$, $A'_x$ and $B'_y$ 
is $P_K P_{A'_x,B'_y|K}$ with 
\begin{multline}
P_{A'_x,B'_y|K}(a,b|k)=P_{\tilde A_x,\tilde B_y}(a,b)\;\mathrm{for}\;k=0 , \\
=P_{F_{x,k}(A_{z(x,k)}),G_{x,k}(B_{t(y,k)})}(a,b)\;\mathrm{for}\;k \ge 1 .
\nonumber
\end{multline}
Summing over $k$ gives the marginal distribution $P_{A'_x,B'_y}=P'_{x,y}$.

For a quantum ${\boldsymbol P}$, the distributions of the $A_x$ 
and $B_y$ can be written as
\begin{equation}
P_{A_x,B_y}(a,b)=\operatorname{tr}(\rho_{\mathrm{lu}} M_{x,a} \otimes N_{y,b}) ,
\label{qp}
\end{equation}
where $\rho_{\mathrm{lu}}$ is a density operator on ${\cal H}_A \otimes {\cal H}_B$ 
and $M_{x,a}$ ($N_{y,b}$) are positive operators such that 
$\sum_{a\in{\cal A}} M_{x,a}=I_A$ ($\sum_{b\in{\cal B}} N_{y,b}=I_B$). 
For any self-maps $F$ on ${\cal A}$ and $G$ on ${\cal B}$, one has
$$P_{F(A_x),G(B_y)}(a,b)=
\sum_{\substack{a'\in F^{-1}(\{ a \})\\b' \in G^{-1}(\{ b \})}} 
P_{A_x,B_y}(a',b') .$$ 
This expression can be recast into the form of eq.\eqref{qp} with $M_{x,a}$ 
and $N_{y,b}$ replaced, respectively, by the operators 
$M_{F,x,a} = \sum_{a' \in F^{-1}(\{ a \})} M_{x,a'}$ and $N_{G,y,b}$ 
defined similarly and so
\begin{multline}
P_{F_{x,k}(A_{z(x,k)}),G_{y,k}(B_{t(y,k)})}(a,b) \\ = \nonumber
\operatorname{tr} (\rho_{\mathrm{lu}} M_{F_{x,k},z(x,k),a} 
\otimes N_{G_{y,k},t(y,k),b} ) .
\end{multline} 

The Bell local distribution tuple $\boldsymbol L$ is also given by 
eq.\eqref{qp} with $\rho_{\mathrm{lu}}$, $M_{x,a}$ and $N_{y,b}$ replaced by 
\begin{equation}
\tilde \rho'_{\mathrm{lu}}=\sum_{k,{\boldsymbol a},{\boldsymbol b}} 
P_K(k) P_{{\boldsymbol {\tilde A}},{\boldsymbol {\tilde B}}}
({\boldsymbol a},{\boldsymbol b}) 
\Pi^{Alice}_{k,{\boldsymbol a},{\boldsymbol b}} \otimes 
\Pi^{Bob}_{k,{\boldsymbol a},{\boldsymbol b}}  ,
\label{tilderho}
\end{equation}
$\sum_{k,{\boldsymbol a},{\boldsymbol b}} \delta_{a_x,a} 
\Pi^{Alice}_{k,{\boldsymbol a},{\boldsymbol b}}$ 
and $\sum_{k,{\boldsymbol a},{\boldsymbol b}} \delta_{b_y,b} 
\Pi^{Bob}_{k,{\boldsymbol a},{\boldsymbol b}}$, respectively, 
where the $\Pi^{Alice}_{k,{\boldsymbol a},{\boldsymbol b}}$ 
($\Pi^{Bob}_{k,{\boldsymbol a},{\boldsymbol b}}$)
are mutually orthogonal rank-one projectors. 
Finally, equation \eqref{qp} with $\rho_{\mathrm{lu}}$, $M_{x,a}$ 
and $N_{y,b}$ replaced by $\tilde \rho'_{\mathrm{lu}} \otimes \rho_{\mathrm{lu}}$,
\begin{equation}
M'_{x,a}=\sum_{k,{\boldsymbol a},{\boldsymbol b}} 
\Pi^{Alice}_{k,{\boldsymbol a},{\boldsymbol b}} 
\otimes \big[\delta_{k,0} \delta_{a_x,a} I_A +   
(1-\delta_{k,0}) M_{F_{x,k},z(x,k),a} \big] ,
\label{Mpxa}
\end{equation}
and $N'_{y,b}$ defined similarly, leads to ${\boldsymbol P}'$, which is 
hence quantum.

\section{Proof of Proposition \ref{R}}\label{PP1}

(i) Let ${\boldsymbol P}$ and ${\boldsymbol P}'$ be two quantum 
distribution tuples with numbers $m$ and $n$ of inputs such that 
the former is not less nonlocal than the latter. For these tuples, 
the Lemma gives the random variables $K$, $\tilde A_x$ and $\tilde B_y$, 
the self-maps $F_{x,k}$ and $G_{y,k}$ and the input maps $z$ and $t$. 
At some stage of any raw key protocol $\mathrm{rkp}$, a random variable $U$ 
($V$) is produced according to $U=A_x$ ($V=B_y$) when $X=x$ ($Y=y$) 
where the $A_x$ ($B_y$) are the random variables among which Alice 
(Bob) can choose and $X$ ($Y$) is a random variable with alphabet in 
$\{ 1, \ldots, m\}$ ($\{ 1, \ldots, n\}$). We name $\mathrm{rkp}_1$ the part of 
$\mathrm{rkp}$ before the generation of $U$ and $V$ and $\mathrm{rkp}_2$ that 
after it. 

From any $\mathrm{rkp}$, we define the protocol $\mathrm{rkp}'$ as follows. First, 
Alice creates $K$, the $\tilde A_x$ and $\tilde B_y$ and sends all of them over the 
public channel. Then, Alice and Bob execute $\mathrm{rkp}_1$ and, instead of 
producing $U$ and $V$ as explained above, they proceed as follows. Alice generates, 
in sequence, $X'$ which is $X$ if $K=0$ and $z(X,k)$ if $K=k \ge 1$, 
$U'$ according to $U'=A_{x'}$ when $X'=x'$ and $U$ which is $\tilde A_x$ if 
$(K,X)=(0,x)$ and $F_{x,k} (U')$ if $(K,X)=(k,x)$ with $k \ge 1$. Similarly, 
Bob generates $Y'$ which is $Y$ if $K=0$ and $t(Y,k)$ if $K=k \ge 1$, 
$V'$ according to $V'=B_{y'}$ when $Y'=y'$ and $V$ which is $\tilde B_y$ if 
$(K,Y)=(0,y)$ and $G_{y,k} (V')$ if $(K,Y)=(k,y)$ with $k \ge 1$. Finally, 
Alice and Bob discard $X'$, $U'$, $Y'$, $V'$, $K$, the $\tilde A_x$ and 
the $\tilde B_y$ which do not play any role in $\mathrm{rkp}_2$ and complete 
$\mathrm{rkp}_2$. One has $S \le R({\boldsymbol P})$ where $S$ is defined 
similarly as $R({\boldsymbol P})$ but taking the supremum only over the raw key 
protocols of the particular form just described. 

Consider any such protocol $\mathrm{rkp}'$, initial tripartite state $\rho$ and 
measurement operators $M_{x,a}$ and $N_{y,b}$ on ${\cal H}_A$ 
and ${\cal H}_B$, respectively, and assume that Alice and Bob perform 
$\mathrm{rkp}'$. After the creation and transmission of $K$, the $\tilde A_x$ and 
the $\tilde B_y$ and the execution of $\mathrm{rkp}_1$, Alice, Bob and Eve share 
the state $\tilde \rho' \otimes \tilde \rho \otimes \rho$ where $\tilde \rho$ 
is given by eq.\eqref{tilderhot} with the distribution 
$P_{{\boldsymbol X},{\boldsymbol Y},{\boldsymbol E}}$ of $\mathrm{rkp}_1$ and
\begin{equation}
\tilde \rho'=\sum_{k,{\boldsymbol a},{\boldsymbol b}} 
P_K(k) P_{{\boldsymbol {\tilde A}},{\boldsymbol {\tilde B}}}
({\boldsymbol a},{\boldsymbol b}) 
\Pi^{Alice}_{k,{\boldsymbol a},{\boldsymbol b}} \otimes 
\Pi^{Bob}_{k,{\boldsymbol a},{\boldsymbol b}} 
\otimes \Pi^{Eve}_{k,{\boldsymbol a},{\boldsymbol b}} , 
\label{tilderhoprime}
\end{equation}
with the same notations as in eq.\eqref{tilderho}. The sum of the projectors 
$\Pi^{Alice}_{k,{\boldsymbol a},{\boldsymbol b}}$ 
($\Pi^{Bob}_{k,{\boldsymbol a},{\boldsymbol b}}$, 
$\Pi^{Eve}_{k,{\boldsymbol a},{\boldsymbol b}}$) is the identity operator 
on a Hilbert space ${\cal H}''_A$ (${\cal H}''_B$, ${\cal H}''_E$). Alice and Bob 
then generate $U$ and $V$ according to $\mathrm{rkp}'$ and discard $X'$, $U'$, 
$Y'$, $V'$, $K$, the $\tilde A_x$ and the $\tilde B_y$, which leads to
\begin{multline}
\omega=
\sum_{{\boldsymbol a},{\boldsymbol b},{\boldsymbol x},{\boldsymbol y}} 
P_{{\boldsymbol {\tilde A}},{\boldsymbol {\tilde B}}}
({\boldsymbol a},{\boldsymbol b})
O_{{\boldsymbol x},{\boldsymbol y}} 
\otimes\Big[ p_0 \Pi^{U,V}_{a_x,b_y} \otimes 
\Pi^{Eve}_{0,{\boldsymbol a},{\boldsymbol b}}
 \otimes \operatorname{tr}_{{\cal H}} \rho \\
+\sum_{k \ge 1,u',v'} p_k \Pi^{U,V}_{F_{x,k}(u'),G_{y,k}(v')} 
 \otimes \Pi^{Eve}_{k,{\boldsymbol a},{\boldsymbol b}} \\
\otimes\operatorname{tr}_{{\cal H}} 
(\rho M_{z(x,k),u'}\otimes N_{t(y,k),v'} \otimes I_E)  \Big] , \nonumber
\end{multline}
where $x$ and $y$ correspond to $X$ and $Y$, respectively, 
$\Pi^{U,V}_{u,v}$ denotes mutually orthogonal rank-one projectors and 
the notations $O_{{\boldsymbol x},{\boldsymbol y}}
=\sum_{{\boldsymbol e}} 
P_{{\boldsymbol X},{\boldsymbol Y},{\boldsymbol E}} 
({\boldsymbol x},{\boldsymbol y},{\boldsymbol e}) 
\Pi_{{\boldsymbol x},{\boldsymbol e}}^{Alice} 
\otimes \Pi_{{\boldsymbol y},{\boldsymbol e}}^{Bob} 
\otimes \Pi_{{\boldsymbol e}}^{Eve}$, $p_k=P_K(k)$ 
and ${\cal H}= {\cal H}_A \otimes {\cal H}_B$ are used. The state $\omega$ 
can be rewritten as
\begin{equation}
\omega=\sum_{{\boldsymbol x},{\boldsymbol y},u,v} 
O_{{\boldsymbol x},{\boldsymbol y}}  
\otimes \Pi_{u,v}^{U,V} \otimes \operatorname{tr}_{{\cal H}'} 
((\tilde \rho' \otimes \rho) (M'_{x,u} \otimes N'_{y,v} \otimes I''_E))  , 
\label{rhopp}
\end{equation}
where $M'_{x,u}$ is given by eq.\eqref{Mpxa}, $N'_{y,v}$ by a similar 
expression, $I''_E$ is the identity operator on 
${\cal H}_E \otimes {\cal H}''_E$ and ${\cal H}'= 
{\cal H}_A \otimes {\cal H}''_A\otimes {\cal H}_B\otimes {\cal H}''_B$. 

As soon as 
$\operatorname{tr}( \rho M_{x,a} \otimes N_{y,b} \otimes I_E)
=P_{x,y}(a,b)$, the state $\tilde \rho' \otimes \rho$, the operators 
$M'_{x,a}$ and $N'_{y,b}$ given by eq.\eqref{Mpxa} and the distributions 
$P'_{x,y}$ are related in the same way, see the proof of the Lemma. 
Thus, one has $R({\boldsymbol P}') \le S'$ where $S'$ is defined 
similarly as $R({\boldsymbol P}')$ but taking the infimum only over such 
particular states and measurement operators. When the state initially shared 
by the three parties is $\tilde \rho' \otimes \rho$ and Alice's and Bob's 
measurements are characterized by the operators $M'_{x,a}$ and 
$N'_{y,b}$, respectively, performing $\mathrm{rkp}_1$ and generating $U$ and $V$ 
according to $\mathrm{rkp}$ gives the state \eqref{rhopp}, see the derivation of 
equation \eqref{rhof}. So, in this case, the tripartite state obtained at the end 
of $\mathrm{rkp}$ is identical to that resulting from the execution of the protocol 
$\mathrm{rkp}'$ with $\rho$, $M_{x,a}$ and $N_{y,b}$. Consequently, $S$ and $S'$ 
are equal to each other, which finishes the proof of property (i).

(ii) Any Bell local distribution tuple can be written in quantum form with 
the state given by eq.\eqref{tilderhoprime} without $K$ and measurement 
operators $\sum_{{\boldsymbol a},{\boldsymbol b}} \delta_{a_x,a} 
\Pi^{Alice}_{{\boldsymbol a},{\boldsymbol b}}$ 
and $\sum_{{\boldsymbol a},{\boldsymbol b}} \delta_{b_y,b} 
\Pi^{Bob}_{{\boldsymbol a},{\boldsymbol b}}$, see the proof of the Lemma. 
Performing any raw key protocol with this initial state and Alice's and Bob's 
measurements described by these operators leads to the state 
\begin{multline}
\rho_{\mathrm{rk}}=
\sum_{\substack{a,b,{\boldsymbol x},{\boldsymbol y},\\{\boldsymbol e},
{\boldsymbol a},{\boldsymbol b}}} 
P_{A|U,{\boldsymbol X},{\boldsymbol E}}
(a|a_x,{\boldsymbol x},{\boldsymbol e})
P_{B|V,{\boldsymbol Y},{\boldsymbol E}}
(b|b_y,{\boldsymbol y},{\boldsymbol e}) \\
\times P_{{\boldsymbol X},{\boldsymbol Y},{\boldsymbol E}}
({\boldsymbol x},{\boldsymbol y},{\boldsymbol e}) 
P_{{\boldsymbol {\tilde A}},{\boldsymbol {\tilde B}}}
({\boldsymbol a},{\boldsymbol b}) 
\Pi_a \otimes \Pi_b \otimes 
\Pi_{\boldsymbol e,{\boldsymbol a},{\boldsymbol b}} ,  \nonumber
\end{multline} 
where $x$ ($y$) corresponds to $X$ ($Y$) and 
$\Pi_{\boldsymbol e,{\boldsymbol a},{\boldsymbol b}}
=\Pi^{Eve}_{\boldsymbol e} \otimes 
\Pi^{Eve}_{{\boldsymbol a},{\boldsymbol b}}$. 
Assume that Eve simply makes the measurement of operators 
$\Pi_{\boldsymbol e,{\boldsymbol a},{\boldsymbol b}}$ on $\rho_{\mathrm{rk}}$. 
The three parties are left with classical random variables. Since 
$P_{{\boldsymbol X},{\boldsymbol Y},{\boldsymbol E}}
=P_{{\boldsymbol X}|{\boldsymbol E}}
P_{{\boldsymbol Y}|{\boldsymbol E}} P_{\boldsymbol E}$, the probability 
mass function of $A$, $B$ and the random variables available to Eve, 
i.e., $\bf {\tilde A}$, $\bf {\tilde B}$ and ${\bf E}$, is 
$P_{A|\bf {\tilde A},{\bf E}} P_{B|\bf {\tilde B},{\bf E}} 
P_{\bf {\tilde A},\bf {\tilde B}} P_{\bf E}$ and hence $A$ and $B$ are 
conditionally independent given Eve's variables. Consequently, Alice and 
Bob cannot generate a secret key \cite{M}.

\section{Proof of Proposition \ref{prop2}}\label{PP2}

For any $\boldsymbol P \in {\cal Q}$, we define the family of distribution 
tuples $\boldsymbol P_p=p\boldsymbol P+(1-p)\boldsymbol L$ 
where $\boldsymbol L$ is any Bell local distribution tuple and $p$ varies 
from $0$ to $1$. They belong to $\cal Q$ since $\cal L \subset \cal Q$ 
and $\cal Q$ is convex \cite{BCPSW}. Clearly, $\boldsymbol P_p$ is 
continuous with respect to $p$, $\boldsymbol P_1=\boldsymbol P$ and 
$\boldsymbol P_0=\boldsymbol L$. Moreover, $\boldsymbol P$ is not 
less nonlocal than $\boldsymbol P_p$ for any $p \in [0,1]$ \cite{dV}. 
We denote by ${\cal Q}_s$ the set of all $\boldsymbol P \in {\cal Q}$ 
such that $N(\boldsymbol P)=s$ and define the function $f$, on the set 
$J'$ of the values of $N$, by 
$f(s)=\inf_{\boldsymbol P \in {\cal Q}_s} M(\boldsymbol P)$. By 
construction, $f \circ N \le M$ on ${\cal Q}$ and there is, for any 
$s \in J'$, $\boldsymbol P \in {\cal Q}_s$ such that $M(\boldsymbol P)$ 
and $f(s)$ are as close to each other as we wish. Since $M$ and $N$ 
vanish on ${\cal L}$, there is a set ${\cal Q}_0$ containing ${\cal L}$ 
and $f(0)=0$. 

As $N \ge 0$, the supremum $N_{\mathrm{sup}}=\sup J'$ is nonnegative. 
Define $J=[0, N_{\mathrm{sup}})$ and consider any $s \in J$. There is 
$\boldsymbol {\hat P} \in {\cal Q}$ such that $N(\boldsymbol {\hat P}) > s$. 
Define $\boldsymbol {\hat P}_p$ as described above. Owing to the continuity 
properties of $N$ and $\boldsymbol {\hat P}_p$, $N(\boldsymbol {\hat P}_p)$ 
is a continuous function of $p$. It is equal to $0$ for $p=0$ and to 
$N(\boldsymbol {\hat P})$ for $p=1$. Thus, due to the intermediate value 
theorem, for any $s' \in [0,N(\boldsymbol {\hat P})]$, there is $q$ such that 
$N(\boldsymbol {\hat P}_q)=s'$, i.e., $s' \in J'$. In particular, $s$ 
belongs to $J'$. As $s$ is any element of $J$, $J$ is a subset of $J'$.

For any $\boldsymbol P \in {\cal Q}_s$, $N(\boldsymbol P_p)$ is a 
continuous function of $p$ which is equal to $0$ for $p=0$ and to $s$ 
for $p=1$. So, for any $s' \in [0,s]$, there is $q$ such that 
$\boldsymbol P_q \in {\cal Q}_{s'}$. Moreover, since $\boldsymbol P$ is 
not less nonlocal than $\boldsymbol P_q$ and $M$ is a nonlocality 
monotone, one has $M(\boldsymbol P)  \ge M(\boldsymbol P_q) \ge f(s')$. 
Thus, for any $s$ and $s'$ in $J'$ such that $s' \le s$, $f(s')$ is a lower 
bound of $M$ on ${\cal Q}_s$, which implies that $f$ is nondecreasing.

\section{Upper bound on the threshold $\tilde N^*$}\label{Ubt}

The rate $R'(\rho_{\mathrm{rk}})$ in eq.\eqref{Rdef} is lowerbounded 
by the Devetak-Winter rate \cite{DW}, i.e.,
$$R'(\rho_{\mathrm{rk}}) \ge 
I({A:B})+\sum\nolimits_a P_A(a) S(\omega_a)-S(\omega) , $$ where 
$I({A:B})$ is the mutual information between $A$ and $B$, $S$ denotes 
the von Neumann entropy, 
$\omega=\operatorname{tr}_{{\cal H}_{AB}} \rho'$ and 
$\omega_a=\operatorname{tr}_{{\cal H}_{AB}} 
(\rho' M_a \otimes I'_B \otimes I'_E)/P_A(a)$ with $I'_B$ the identity 
operator on ${\cal H}_B \otimes {\cal H}'_B$. We 
consider $m=n=2$, random variables $A_x$ and $B_y$ with values in 
$\{-1,1\}$ and a raw key protocol in which Alice creates three equally 
distributed random variables, $X$, $Y$ and $E$ and sends $Y$ and $E$ 
over the public channel, $A=\nu(X,Y)EU$ where $\nu$ is a map from 
$\{ 1,2\}^2$ to $\{ -1,1\}$ such that $\nu(x,y)=-1$ for only one pair 
$(x,y)$ and $B=EV$. The values of $E$ are $-1$ and $1$, $X$ and $Y$ 
are the choice random variables for Alice and Bob, respectively, with 
alphabet $\{ 1,2\}$. Consequently, $P_A=P_B=1/2$ and the above Eve's 
states are given by $\omega = \sum_{y,e} \Pi_{y,e} 
\otimes \operatorname{tr}_{{\cal H}_A \otimes {\cal H}_B} \rho/4$ 
and $$\omega_a =\sum_{x,y,e} \Pi_{y,e} \otimes 
\operatorname{tr}_{{\cal H}_A \otimes {\cal H}_B} 
(\rho M_{x,e\nu(x,y)a} \otimes I_B \otimes I_E)/4 , $$
omitting the superscript for the projectors. 

Since $m=n=2$ and ${\cal A}={\cal B}=\{-1,1 \}$, these states can be 
rewritten as $\omega = \sum_{y,e,\lambda}  p_\lambda \Pi_{y,e} 
\otimes\operatorname{tr}_{{\cal H}_2^2} \rho_\lambda/4$ and
$$\omega_a = \sum_{x,y,e,\lambda} \frac{p_\lambda}{8}
\Pi_{y,e} \otimes \operatorname{tr}_{{\cal H}_2^2} 
(\rho_\lambda (I_2+e\nu a \Sigma^A_{x,\lambda}) 
\otimes I_2 \otimes I_E) , $$ omitting the arguments of $\nu$, 
where $p_\lambda$ denotes probabilities summing to unity, ${\cal H}_2$ 
the Hilbert space of dimension 2, $\rho_\lambda$ density operators on 
${\cal H}_2^2\otimes{\cal H}_E$, and $I_2$ the identity operator on ${\cal H}_2$. 
In some basis of ${\cal H}_2$, depending on $\lambda$, the diagonal elements 
of the operators $\Sigma^A_{x,\lambda}$ can be expressed as 
$\pm \cos \theta_{x,\lambda}$ and the nondiagonal ones as 
$\sin \theta_{x,\lambda}$ \cite{PABGMS}. In terms of the states 
$\rho_\lambda$, the distributions $P_{x,y}$ read as 
$$P_{x,y}(a,b)=\sum_\lambda \frac{p_\lambda}{4}
\operatorname{tr}( \rho_\lambda  (I_2+a \Sigma^A_{x,\lambda}) \otimes 
(I_2+b \Sigma^B_{y,\lambda}) \otimes I_E) , $$ 
where the operators $\Sigma^B_{y,\lambda}$ are similar to the 
$\Sigma^A_{x,\lambda}$. 

The above Eve's states can be further simplified into
$\omega = \sum_\lambda  p_\lambda 
\operatorname{tr}_{{\cal H}_2^2} \rho'_\lambda$ and
$$\omega_a = \sum_{x,\lambda} \frac{p_\lambda}{2}
 \operatorname{tr}_{{\cal H}_2^2} 
( \rho'_\lambda (I_2+a \Sigma^A_{x,\lambda}) \otimes I_2 \otimes I'_E) . $$
The states $\rho'_\lambda$ are given by 
$\rho'_\lambda=\sum_{y,e} \Pi_{y,e} \otimes \rho_{\lambda,e\nu}/4$ 
where $\rho_{\lambda,1}=\rho_{\lambda}$ and 
$\rho_{\lambda,-1}=\sigma^A_{\lambda} \otimes \sigma^B_{\lambda} 
\otimes I_E \rho_{\lambda} \sigma^A_{\lambda} 
\otimes \sigma^B_{\lambda} \otimes I_E$ 
with $\sigma^A_{\lambda}=(\substack{0 \, -i\\ i \;\;\; 0})$ in the basis in 
which the $\Sigma^A_{x,\lambda}$ are real and similarly for 
$\sigma^B_{\lambda}$. The above expression for $\omega_a$ results 
from $\sigma^A_{\lambda}\Sigma^A_{x,\lambda} \sigma^A_{\lambda}
=-\Sigma^A_{x,\lambda}$. The reduced density operator on 
${\cal H}_2^2$ of $\rho'_\lambda$ is 
$\operatorname{tr}_{{\cal H}_E} (\rho_{\lambda,1}+\rho_{\lambda,-1})/2$ 
which can always be taken to be a Bell diagonal state and hence
$$S(\omega_\lambda)-\sum_a S(\omega_{\lambda,x,a})/2
\le h([1+(N_\lambda+N_\lambda^2/4)^{1/2}]/2) , $$ 
where $\omega_\lambda=\operatorname{tr}_{{\cal H}_2^2} \rho'_\lambda$, 
$\omega_{\lambda,x,a}=\operatorname{tr}_{{\cal H}_2^2} 
( \rho'_\lambda (I+a \Sigma^A_{x,\lambda}) \otimes I \otimes I'_E)$, 
$h$ is the binary entropy function and $N_\lambda$ is the maximum violation 
of the Clauser-Horne-Shimony-Holt inequality \cite{CHSH} for the state 
$\operatorname{tr}_{{\cal H}_E} (\rho_{\lambda,1}+\rho_{\lambda,-1})/2$ 
\cite{PABGMS}. As $\sum_{a,b} ab P_{x,y}(a,b)=\sum_{a,b} ab P_{x,y}(-a,-b)$, 
the value of $\langle A_x B_y \rangle$ remains the same when 
$\rho_\lambda$ is replaced by $(\rho_{\lambda,1}+\rho_{\lambda,-1})/2$ 
and so $\tilde N({\bf P}) \le \sum_\lambda p_\lambda N_\lambda$ where 
$\tilde N$ is defined by eq.\eqref{CHSH}. Thus, due to the properties of the Holevo 
quantity and of $h$, the above inequality is valid with $\omega_\lambda$, 
$\omega_{\lambda,x,a}$ and $N_\lambda$ replaced, respectively, by $\omega$, 
$\omega_a$ and $\tilde N({\bf P})$.

Since $P_A=P_B=1/2$, one has $I({A:B})=1 - h(1/2+|\langle A B \rangle|/2)$ 
where $$\langle A B \rangle=\langle \nu(X,Y) UV \rangle
=\frac{1}{4}\sum_{x,y} \nu(x,y) \langle A_x B_y \rangle , $$
and thus $\max_\nu I({A:B}) \ge 1 - h(3/4+\tilde N({\bf P})/8)$. 
The above results show that $R \ge g' \circ \tilde N$ where $g'$ is given by 
$$g'(s)=1-h\left(\frac{3}{4}+\frac{s}{8}\right)
-h\left(\frac{1}{2}+\frac{1}{2}\sqrt{s+\frac{s^2}{4}}\right) . $$ 
As $R$ is nonnegative, the right side of the above inequality can be replaced by 
zero when it is negative and hence $R \ge g \circ \tilde N$ with 
$g(s)=\max\{0,g'(s)\}$. The value $g'(s)$ is positive for $s\ge 0.652$. So, 
the nonlocality monotone $\tilde N$ has a threshold value 
$\tilde N^*\le 0.652$.

\end{appendix}

\end{document}